\documentclass{aastex631}
\usepackage[T5,T1]{fontenc}
\usepackage{natbib}
\usepackage{url}
\usepackage{xcolor}
\usepackage{graphicx}
\usepackage{booktabs}
\usepackage{multirow}
\usepackage{amsmath}

\graphicspath{{figures/}}

\newcommand{\txs}{TXS~0506+056}
\newcommand{\ngc}{NGC~1068}

\definecolor{darkgreen}{rgb}{0.0, 0.5, 0.0}

\begin{document}

\title{IceCube Second Track Data Release IceTracks-DR2: Data from 2008-2022 for Neutrino Source Searches}

\affiliation{III. Physikalisches Institut, RWTH Aachen University, D-52056 Aachen, Germany}
\affiliation{Department of Physics, University of Adelaide, Adelaide, 5005, Australia}
\affiliation{Dept. of Physics and Astronomy, University of Alaska Anchorage, 3211 Providence Dr., Anchorage, AK 99508, USA}
\affiliation{School of Physics and Center for Relativistic Astrophysics, Georgia Institute of Technology, Atlanta, GA 30332, USA}
\affiliation{Dept. of Physics, Southern University, Baton Rouge, LA 70813, USA}
\affiliation{Dept. of Physics, University of California, Berkeley, CA 94720, USA}
\affiliation{Lawrence Berkeley National Laboratory, Berkeley, CA 94720, USA}
\affiliation{Institut f{\"u}r Physik, Humboldt-Universit{\"a}t zu Berlin, D-12489 Berlin, Germany}
\affiliation{Fakult{\"a}t f{\"u}r Physik {\&} Astronomie, Ruhr-Universit{\"a}t Bochum, D-44780 Bochum, Germany}
\affiliation{Universit{\'e} Libre de Bruxelles, Science Faculty CP230, B-1050 Brussels, Belgium}
\affiliation{Vrije Universiteit Brussel (VUB), Dienst ELEM, B-1050 Brussels, Belgium}
\affiliation{Dept. of Physics, Simon Fraser University, Burnaby, BC V5A 1S6, Canada}
\affiliation{Department of Physics and Laboratory for Particle Physics and Cosmology, Harvard University, Cambridge, MA 02138, USA}
\affiliation{Dept. of Physics, Massachusetts Institute of Technology, Cambridge, MA 02139, USA}
\affiliation{Dept. of Physics and The International Center for Hadron Astrophysics, Chiba University, Chiba 263-8522, Japan}
\affiliation{Department of Physics, Loyola University Chicago, Chicago, IL 60660, USA}
\affiliation{Dept. of Physics and Astronomy, University of Canterbury, Private Bag 4800, Christchurch, New Zealand}
\affiliation{Dept. of Physics, University of Maryland, College Park, MD 20742, USA}
\affiliation{Dept. of Astronomy, Ohio State University, Columbus, OH 43210, USA}
\affiliation{Dept. of Physics and Center for Cosmology and Astro-Particle Physics, Ohio State University, Columbus, OH 43210, USA}
\affiliation{Niels Bohr Institute, University of Copenhagen, DK-2100 Copenhagen, Denmark}
\affiliation{Dept. of Physics, TU Dortmund University, D-44221 Dortmund, Germany}
\affiliation{Dept. of Physics and Astronomy, Michigan State University, East Lansing, MI 48824, USA}
\affiliation{Dept. of Physics, University of Alberta, Edmonton, Alberta, T6G 2E1, Canada}
\affiliation{Erlangen Centre for Astroparticle Physics, Friedrich-Alexander-Universit{\"a}t Erlangen-N{\"u}rnberg, D-91058 Erlangen, Germany}
\affiliation{Physik-department, Technische Universit{\"a}t M{\"u}nchen, D-85748 Garching, Germany}
\affiliation{D{\'e}partement de physique nucl{\'e}aire et corpusculaire, Universit{\'e} de Gen{\`e}ve, CH-1211 Gen{\`e}ve, Switzerland}
\affiliation{Dept. of Physics and Astronomy, University of Gent, B-9000 Gent, Belgium}
\affiliation{Dept. of Physics and Astronomy, University of California, Irvine, CA 92697, USA}
\affiliation{Karlsruhe Institute of Technology, Institute for Astroparticle Physics, D-76021 Karlsruhe, Germany}
\affiliation{Karlsruhe Institute of Technology, Institute of Experimental Particle Physics, D-76021 Karlsruhe, Germany}
\affiliation{Dept. of Physics, Engineering Physics, and Astronomy, Queen's University, Kingston, ON K7L 3N6, Canada}
\affiliation{Department of Physics {\&} Astronomy, University of Nevada, Las Vegas, NV 89154, USA}
\affiliation{Nevada Center for Astrophysics, University of Nevada, Las Vegas, NV 89154, USA}
\affiliation{Dept. of Physics and Astronomy, University of Kansas, Lawrence, KS 66045, USA}
\affiliation{UCLouvain, Centre for Cosmology, Particle Physics and Phenomenology, CP3, Chemin du Cyclotron 2, 1348 Louvain-la-Neuve, Belgium}
\affiliation{Department of Physics, Mercer University, Macon, GA 31207-0001, USA}
\affiliation{Dept. of Astronomy, University of Wisconsin{\textemdash}Madison, Madison, WI 53706, USA}
\affiliation{Dept. of Physics and Wisconsin IceCube Particle Astrophysics Center, University of Wisconsin{\textemdash}Madison, Madison, WI 53706, USA}
\affiliation{Institute of Physics, University of Mainz, Staudinger Weg 7, D-55099 Mainz, Germany}
\affiliation{Department of Physics, Marquette University, Milwaukee, WI 53201, USA}
\affiliation{Institut f{\"u}r Kernphysik, Universit{\"a}t M{\"u}nster, D-48149 M{\"u}nster, Germany}
\affiliation{Bartol Research Institute and Dept. of Physics and Astronomy, University of Delaware, Newark, DE 19716, USA}
\affiliation{Dept. of Physics, Yale University, New Haven, CT 06520, USA}
\affiliation{Columbia Astrophysics and Nevis Laboratories, Columbia University, New York, NY 10027, USA}
\affiliation{Dept. of Physics, University of Oxford, Parks Road, Oxford OX1 3PU, United Kingdom}
\affiliation{Dipartimento di Fisica e Astronomia Galileo Galilei, Universit{\`a} Degli Studi di Padova, I-35122 Padova PD, Italy}
\affiliation{Dept. of Physics, Drexel University, 3141 Chestnut Street, Philadelphia, PA 19104, USA}
\affiliation{Physics Department, South Dakota School of Mines and Technology, Rapid City, SD 57701, USA}
\affiliation{Dept. of Physics, University of Wisconsin, River Falls, WI 54022, USA}
\affiliation{Dept. of Physics and Astronomy, University of Rochester, Rochester, NY 14627, USA}
\affiliation{Department of Physics and Astronomy, University of Utah, Salt Lake City, UT 84112, USA}
\affiliation{Dept. of Physics, Chung-Ang University, Seoul 06974, Republic of Korea}
\affiliation{Oskar Klein Centre and Dept. of Physics, Stockholm University, SE-10691 Stockholm, Sweden}
\affiliation{Dept. of Physics and Astronomy, Stony Brook University, Stony Brook, NY 11794-3800, USA}
\affiliation{Dept. of Physics, Sungkyunkwan University, Suwon 16419, Republic of Korea}
\affiliation{Institute of Physics, Academia Sinica, Taipei, 11529, Taiwan}
\affiliation{Dept. of Physics and Astronomy, University of Alabama, Tuscaloosa, AL 35487, USA}
\affiliation{Dept. of Astronomy and Astrophysics, Pennsylvania State University, University Park, PA 16802, USA}
\affiliation{Dept. of Physics, Pennsylvania State University, University Park, PA 16802, USA}
\affiliation{Dept. of Physics and Astronomy, Uppsala University, Box 516, SE-75120 Uppsala, Sweden}
\affiliation{Dept. of Physics, University of Wuppertal, D-42119 Wuppertal, Germany}
\affiliation{Deutsches Elektronen-Synchrotron DESY, Platanenallee 6, D-15738 Zeuthen, Germany}

\author[0000-0001-6141-4205]{R. Abbasi}
\affiliation{Department of Physics, Loyola University Chicago, Chicago, IL 60660, USA}

\author[0000-0001-8952-588X]{M. Ackermann}
\affiliation{Deutsches Elektronen-Synchrotron DESY, Platanenallee 6, D-15738 Zeuthen, Germany}

\author{J. Adams}
\affiliation{Dept. of Physics and Astronomy, University of Canterbury, Private Bag 4800, Christchurch, New Zealand}

\author[0000-0003-2252-9514]{J. A. Aguilar}
\affiliation{Universit{\'e} Libre de Bruxelles, Science Faculty CP230, B-1050 Brussels, Belgium}

\author[0000-0003-0709-5631]{M. Ahlers}
\affiliation{Niels Bohr Institute, University of Copenhagen, DK-2100 Copenhagen, Denmark}

\author[0000-0002-9534-9189]{J.M. Alameddine}
\affiliation{Dept. of Physics, TU Dortmund University, D-44221 Dortmund, Germany}

\author[0009-0001-2444-4162]{S. Ali}
\affiliation{Dept. of Physics and Astronomy, University of Kansas, Lawrence, KS 66045, USA}

\author{N. M. Amin}
\affiliation{Bartol Research Institute and Dept. of Physics and Astronomy, University of Delaware, Newark, DE 19716, USA}

\author[0000-0001-9394-0007]{K. Andeen}
\affiliation{Department of Physics, Marquette University, Milwaukee, WI 53201, USA}

\author[0000-0003-4186-4182]{C. Arg{\"u}elles}
\affiliation{Department of Physics and Laboratory for Particle Physics and Cosmology, Harvard University, Cambridge, MA 02138, USA}

% ...

\date{\today}

\collaboration{415}{IceCube Collaboration}

\email{analysis@icecube.wisc.edu}

\begin{abstract}
We present IceCube's latest release of muon track data for neutrino point-source searches, extending the previously published 10-year dataset to cover 14 years of observations (April 6, 2008 – May 23, 2022). This release features an updated event selection and improved detector calibration for data recorded after June 1, 2010. The release also includes binned instrument response functions and effective areas, enabling the community to perform sensitive searches for steady and transient neutrino sources. We report on key science results obtained with this dataset using internal IceCube analysis tools and compare them to those derived from analyses based on the binned response functions included in this public release. To facilitate reproducible research, we provide benchmark results obtained using this data release and publicly available software. This release represents IceCube's most sensitive and comprehensive publicly available all-sky muon track dataset to date and should be preferred over previous releases.
\end{abstract}


\section{Introduction}

The IceCube Neutrino Observatory is a cubic-kilometer water Cherenkov detector embedded in the Antarctic ice at the South Pole. Designed to identify neutrinos ranging in energy from GeV to PeV, IceCube has measured a diffuse astrophysical neutrino flux~\citep{IceCube:2013Science} and identified candidate point sources, including \txs~\citep{IceCube:2018Science_alert,IceCube:2018Science_flare} and \ngc~\citep{IceCube:2022Science}, while also detecting a neutrino flux from the Milky Way~\citep{IceCube:2023Science}.

The in-ice component of IceCube is composed of 5160 Digital Optical Modules (DOMs) distributed over a cubic kilometer of ice between depths of 1.45 km and 2.45 km~\citep{2017JInst..12P3012A}. Light emitted by charged particles traveling through the ice can be detected using these modules, allowing for the reconstruction of the primary particle energy, direction, and type.  In this way, IceCube detects various types of primary particles, including astrophysical neutrinos, atmospheric neutrinos, and atmospheric muons, which manifest in a variety of event topologies classified into the broad categories of tracks and cascades.

The distribution of observed charge in the IceCube detector differs for different particle interactions
in the ice. For example, muons, generated either in the interactions of muon neutrinos or in air showers created by cosmic-ray interactions in the atmosphere, produce tracks with elongated light patterns in the detector. Thanks to their elongated light patterns, tracks allow for precise directional reconstruction, with a median angular separation between the parent neutrino and the reconstructed muon below 1$^{\circ}$.  Tracks are thus especially valuable in searches for astrophysical neutrino point sources. In contrast, cascade events, typically generated by electron neutrinos, have a much more compact geometry and are less suitable for searching for point sources, though have notably been used to observe emission from the galactic plane~\citep{IceCube:2023Science}. The data presented here (IceTracks-DR2) focuses on the tracks channel, and is intended to aid further studies in multi-messenger astrophysics. 

To support a broad range of scientific studies and to facilitate collaboration across the multi-messenger astrophysics community, IceCube has previously released high-quality final-level data samples of neutrino candidate events. Past releases have been provided in various formats to serve both scientific and technical uses\footnote{\url{https://icecube.wisc.edu/science/data-releases/}}, enabling members of the community to perform independent analyses, develop new methods, and contribute to the interpretation of astrophysical neutrino observations. Building on this effort, we present an updated dataset of track-like events selected as muon-neutrino candidates, extending the previously publicly available 2021 data release (hereafter IceTracks-DR1, \citealt{IceCube:2021arXiv}) to cover a total of 14 years. This new release, which we refer to as IceTracks-DR2, includes four additional years of data collected with the full IceCube detector configuration and incorporates improved calibrations~\citep{Icecube:2021PhRvD} compared to earlier versions. The dataset is accompanied by relevant detector response information to enable accurate interpretation and modeling. This paper documents the contents of the release, outlines the key updates and improvements compared to IceTracks-DR1~\citep{IceCube:2021arXiv}, and provides validation through the reproduction of selected published results.

\section{Event Selection}\label{sec:evt-selection}
The event selection presented as part of this data release is a high-statistics sample of track-like neutrino event candidates observed by IceCube between 2008 and 2021. This includes both atmospheric and astrophysical neutrinos, as well as a population of high-energy atmospheric muons in the Southern sky. Data taking periods are split into ``seasons", denoted as ``IC\#\#", corresponding to the construction status of the detector during a given period. IceTracks-DR2 contains 4 detector configurations: IC40, IC59, IC79, and IC86. The specific start and end dates, as well as the total number of events observed during each year of data collection, can be seen in Table~\ref{tab:seasonlist}. Note that in contrast to previous data releases, all IC86 seasons are now subject to the same event processing and detector response. There is no longer a distinction between IC86-2011 and the other IC86 seasons. 

The event selection begins with a set of filters that selects for events with muon tracks from all directions that could be induced by neutrinos. As these events are identified only by the amount of charge  and reconstructed zenith angle, the sample contains a large number of atmospheric muons. At this stage in the selection process the recorded rate of events is approximately 40 Hz.

Additional event reconstructions are then performed and events are evaluated according to their reconstructed zenith angle. For the purposes of this event selection, we consider the ``Northern Hemisphere" to be defined as declinations $\ge$ -5 degrees, as Earth and the Antarctic glacier sufficiently shield the cosmic ray background up to this point. Events in the Northern hemisphere (declination $\ge -5$ degrees) are selected based on event quality properties including the number of DOMs illuminated and the length of the reconstructed track. Events in the Southern hemisphere (declination $< -5$ degrees) are subject to similar requirements as events in the Northern hemisphere, with some additional and modified cuts based on the number of IceCube strings illuminated and the reconstructed angular error. The sample also includes an effective cut on the minimum event energy in the southern sky, as astrophysical neutrino fluxes are expected to have a harder energy spectrum than atmospheric muons and neutrinos. The southern-sky selection is tuned to produce a similar event rate as the northern sky event selection, resulting in a final all-sky event rate of approximately 4 mHz. Events from both hemispheres are then evaluated by a boosted decision tree (BDT) trained to select for track-like events while rejecting cascade-like events. 

\begin{figure}
    \centering
    \includegraphics[width=0.4\linewidth]{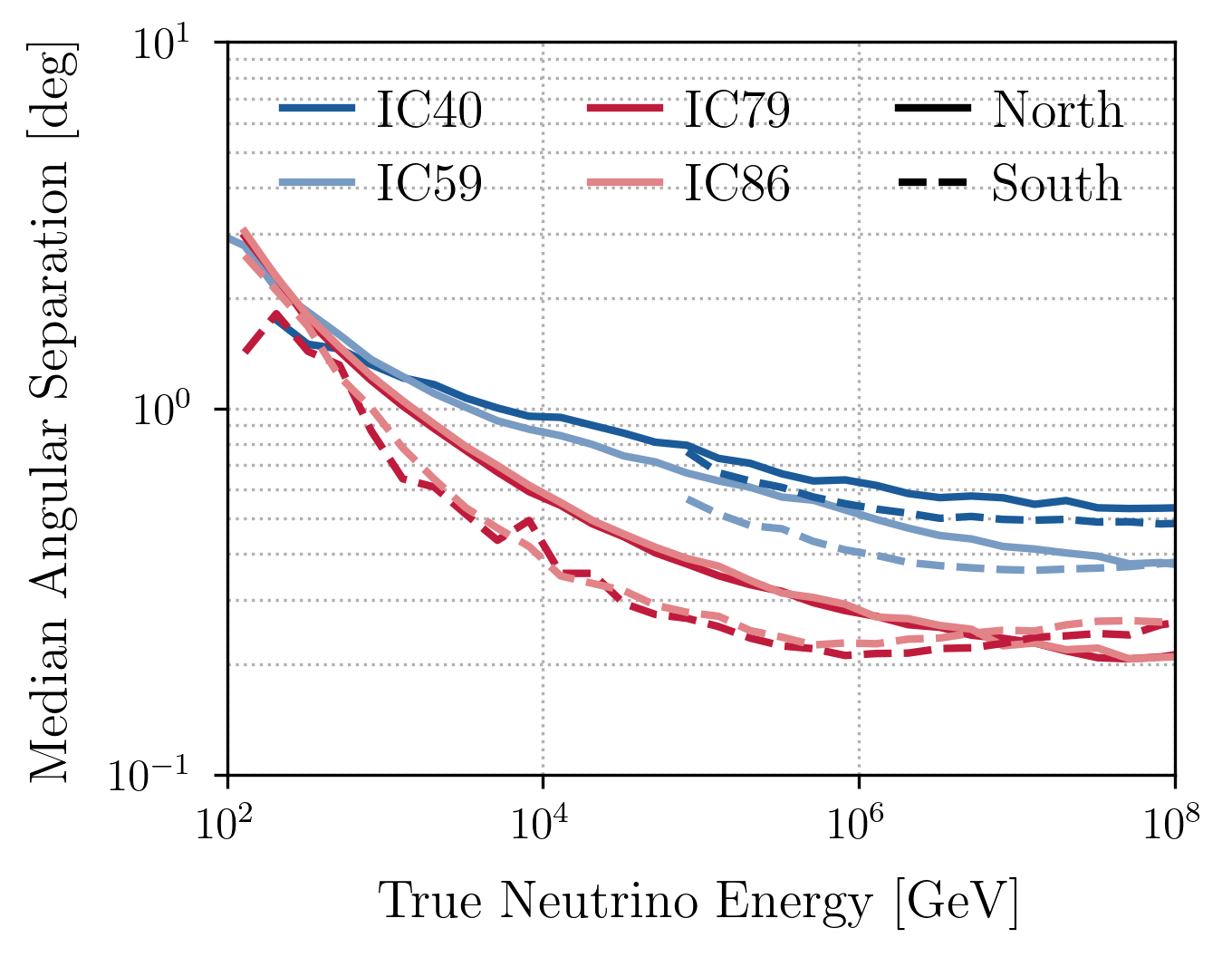}
    \caption{Median angular separation between the simulated neutrino and the reconstructed muon directions as a function of the true neutrino energy. Different colors represent different detector seasons. Solid and dashed lines are for northern and southern sky, respectively. For IC40 and IC59, the southern sky median is not shown at low energies because too few events pass the selection in that region to yield a stable estimate.}
    \label{fig:ang-reso}
\end{figure}

After all selection criteria are applied, the resulting sample exhibits a median angular separation between the parent neutrino and the reconstructed muon reaching below 1 degree at energies above 10~TeV, both in the northern and southern hemispheres (see Figure~\ref{fig:ang-reso}). The complete or almost complete detector configurations with 86 and 79 strings, respectively, outperform the smaller detector seasons due to the increased amount of instrumentation available.

Data from the IC79 to IC86-2018 seasons has been reprocessed with updated detector calibration and online event filtering. This update is commonly referred to as ``Pass2" processing. As such, the event content of the pre-2018 seasons has changed relative to previous data releases, reflecting the updated detector calibration that improves energy resolution, angular resolution, and event classification. Table~\ref{tab:overlap_numbers} lists the per-season overlap of IceTracks-DR2 with the previous IceTracks-DR1. The IC79 and IC86-2011 seasons are most affected, with the events in IceTracks-DR2 having less than 50\% overlap with IceTracks-DR1 during these seasons. The difference is due to updates to the handling of coincident events and improved reconstructions changing the evaluation of lower energy southern sky events, as this region appears to display the most dramatic difference in event content: IC86-2011 has 77\% overlap between the old and new data release in the northern sky, but only 12.8\% in the southern sky. 

\begin{table}[htbp]
     \centering
     \setlength{\tabcolsep}{18pt}
     \begin{tabular}{c  c  c  c  c  c} 
     \toprule
       Season & Start & End & Livetime & Events & Ref \\
     \midrule
     \midrule
     IC40 & 2008/04/06 & 2009/05/20 & 376.4 & 36900 & \citet{Abbasi:2010rd}\\  
     IC59 & 2009/05/20 & 2010/05/31 & 353.6 & 107011  & \citet{Aartsen:2013uuv}\\ 
     IC79 & 2010/06/01 & 2011/05/13 & 312.8 & 101956  & \citet{Aartsen:2019fau}\\
     IC86-2011 & 2011/05/13 & 2012/05/15 & 339.0 & 118774  & \citet{Aartsen:2019fau}\\
     IC86-2012 & 2012/04/26 & 2013/05/02 & 327.1 & 116669  & \citet{Aartsen:2019fau}\\
     IC86-2013 & 2013/04/18 & 2014/05/06 & 356.2 & 126535  & \citet{Aartsen:2019fau}\\
     IC86-2014 & 2014/04/10 & 2015/05/18 & 364.9 & 129520  & \citet{Aartsen:2019fau}\\
     IC86-2015 & 2015/04/24 & 2016/05/20 & 365.3 & 130389  & \citet{Aartsen:2019fau}\\
     IC86-2016 & 2016/05/20 & 2017/05/18 & 357.2 & 126260  & \citet{Aartsen:2019fau}\\
     IC86-2017 & 2017/05/18 & 2018/07/10 & 410.9 & 147460  & \citet{Aartsen:2019fau}\\
     IC86-2018 & 2018/06/19 & 2019/07/17 & 368.8 & 131516  & \citet{Aartsen:2019fau}\\
     IC86-2019 & 2019/06/28 & 2020/05/29 & 313.3 & 112830  & \citet{Aartsen:2019fau}\\
     IC86-2020 & 2020/05/08 & 2021/05/27 & 361.3 & 129840  & \citet{Aartsen:2019fau}\\
     IC86-2021 & 2021/05/03 & 2022/05/23 & 356.6 & 127695  & \citet{Aartsen:2019fau}\\
     \bottomrule
    \end{tabular}
    \caption{\textbf{Event Sample Information.} Event samples by IceCube season, including season start/end times, livetimes, and event counts. References for detailed descriptions of the sample selection for each detector configuration are also included. Seasons may overlap, as each detector season typically begins with several weeks of test processing before fully completing the transition. Note that in DR2, the IC79 and IC86-2011 season event processing has been unified with the other IC86 seasons. This has been changed from DR1, where both seasons had unique processing procedures.}
    \label{tab:seasonlist}
\end{table}

\begin{table}[htbp]
    \centering
    \setlength{\tabcolsep}{29pt}
    \begin{tabular}{c  c  c  c} 
     \toprule
       Season & DR1 Event Count & \% of DR1 in DR2 &  \% of DR2 in DR1  \\
     \midrule
     \midrule
     IC40 & 36900 & 100\% & 100\% \\  
     IC59 & 107011 & 100\% & 100\% \\ 
     IC79 & 93133 & 49.8\% & 45.5\% \\
     IC86-2011 & 136244 & 39.5\% & 45.3\% \\
     IC86-2012 & 112858 & 90.3\% & 87.3\% \\
     IC86-2013 & 122541 & 89.1\% & 86.3\% \\
     IC86-2014 & 127045 & 89.7\% & 88.0\% \\
     IC86-2015 & 129311 & 97.1\% & 96.3\% \\
     IC86-2016 & 123657 & 94.1\% & 92.2\% \\
     IC86-2017 & 145750 & 99.9\% & 98.8\% \\ 
     \bottomrule
    \end{tabular}
    \caption{\textbf{Percentage of Events in Each Season.} The percentage of events that exist in both the new IceTracks-DR2 and the previous IceTracks-DR1 data releases by season. Seasons after IC86-2018 are unique to IceTracks-DR2. Differences are due to changes in event processing and detector calibration. Notably, these affect the IC79--IC86-2011 seasons more as in IceTracks-DR1 these seasons had previously used processing unique to that season, which has now been updated to be standardized with the rest of the IC86 seasons.}
    \label{tab:overlap_numbers}
\end{table}

\section{Detector Response}\label{sec:detector-response}
Detailed simulation is used to evaluate the response of IceCube to the various types of events seen by the detector. For the purposes of this data release, the relevant simulations can be characterized by the detector effective area ($A_\mathrm{eff}$) and a set of functions describing the detector response, characterized as a 5-dimensional response matrix (the ``Instrument Response Function" or ``IRF" matrix) spanned by neutrino energy, neutrino declination, reconstructed event (muon) energy, point spread function, and estimated angular uncertainty. 

\begin{figure}
  \centering
  \includegraphics[width=\textwidth]{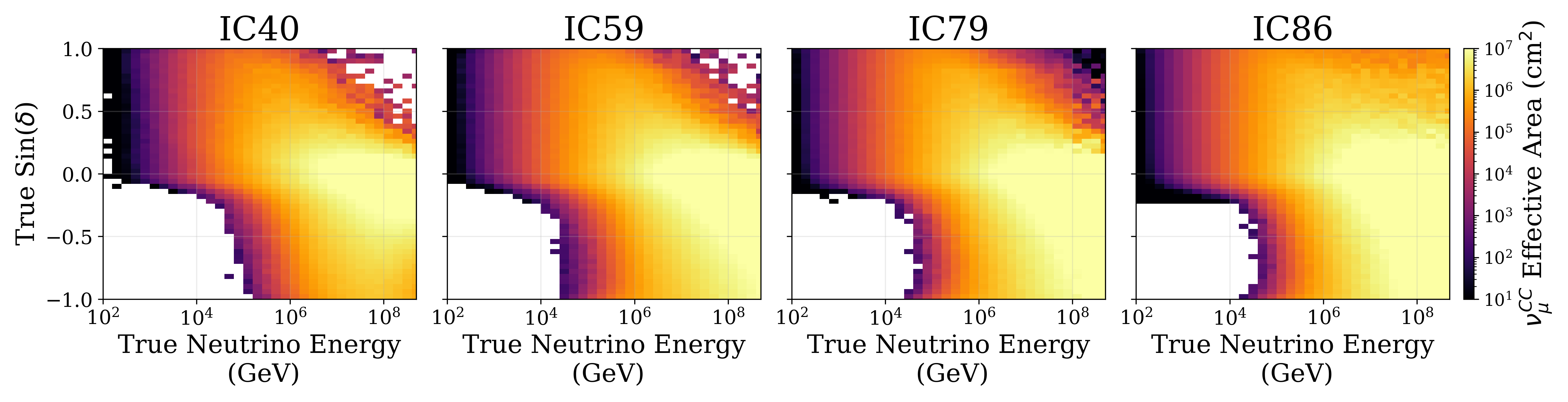}
  \caption{The binned effective areas for each data-taking period included in IceTracks-DR2. IC40 through IC79 use the partially-completed detector while IC86 uses the complete 86-string detector configuration.}
  \label{fig:effA}
\end{figure}

The number of expected neutrino events observed by the detector can be calculated as:

\begin{align}\label{eq:ev_rate}
N_\nu=\int dt \int d\Omega\int_0^\infty dE_{\nu}\,A_\mathrm{eff}\left(E_{\nu}, \Omega\right) \phi_\nu\left(E_\nu,\Omega, t\right),
\end{align}

\noindent where $\phi_\nu\left(E_\nu,\Omega, t\right)$ denotes the incident neutrino flux. The effective area $A_\mathrm{eff}$ varies with neutrino energy and declination, and differs with the detector configuration. Plots of the IceCube effective area for each detector configuration can be seen in Figure~\ref{fig:effA}. This effective area is largely similar to that which was provided with IceTracks-DR1, however it has been improved in a few key ways. The effective areas included in IceTracks-DR2 are generated with new ``Pass2" simulation incorporating improvements to detector calibration and filtering. Additionally, low-energy misreconstructed events in the southern sky have been manually removed from the reported effective areas, as these events correspond to rare triggers on coincident simulated background events, leading to an overestimation of the effective area in specific energy and angular bins. See the discussion in section \ref{sec:benchmarkresults} for further information on how this affects generic point source analyses constructed using publicly available materials. 

The Instrument Response Function matrix allows for modeling of observed event properties for simulated events with known true energy and direction. The IRF matrix bins are chosen to be as small as possible while still reproducing roughly equivalent sensitivity and observed limits for generic point source searches as the more detailed internal parameterizations used within the IceCube Collaboration. Unlike IceTracks-DR1, IRF matrix declination bins are chosen to match binning used to build PDFs for IceCube's published analyses, allowing for a finer description of the instrument response as a function of incident event energy and direction. Note that IRF matrix bins are not constructed to span equal declinations, rather bins are defined to be smaller in declination near the horizon, where the detector response varies most rapidly as a function of the declination of the incoming event. 

\begin{figure}
  \centering
  \includegraphics[width=0.32\textwidth]{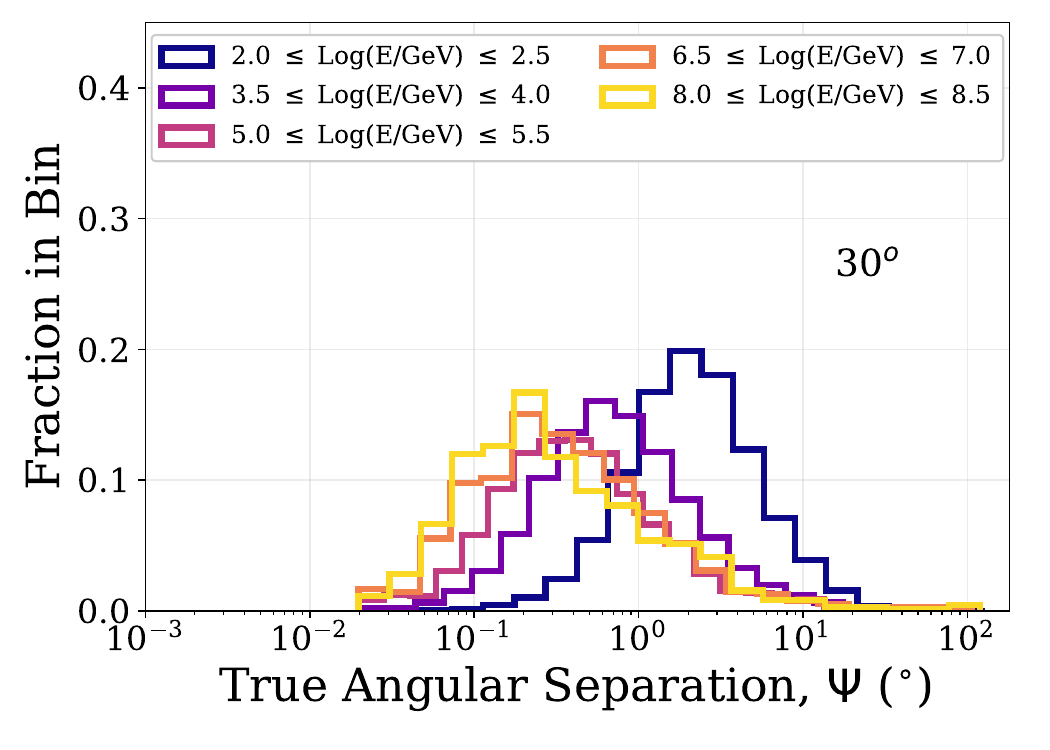}
  \includegraphics[width=0.32\textwidth]{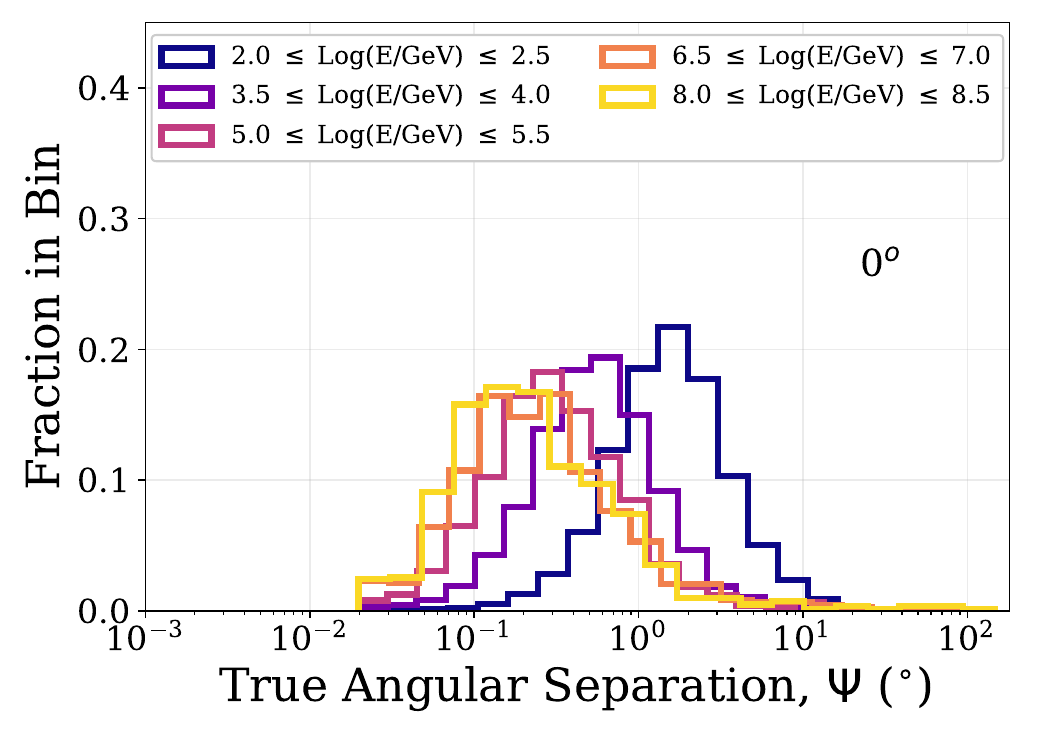}
  \includegraphics[width=0.32\textwidth]{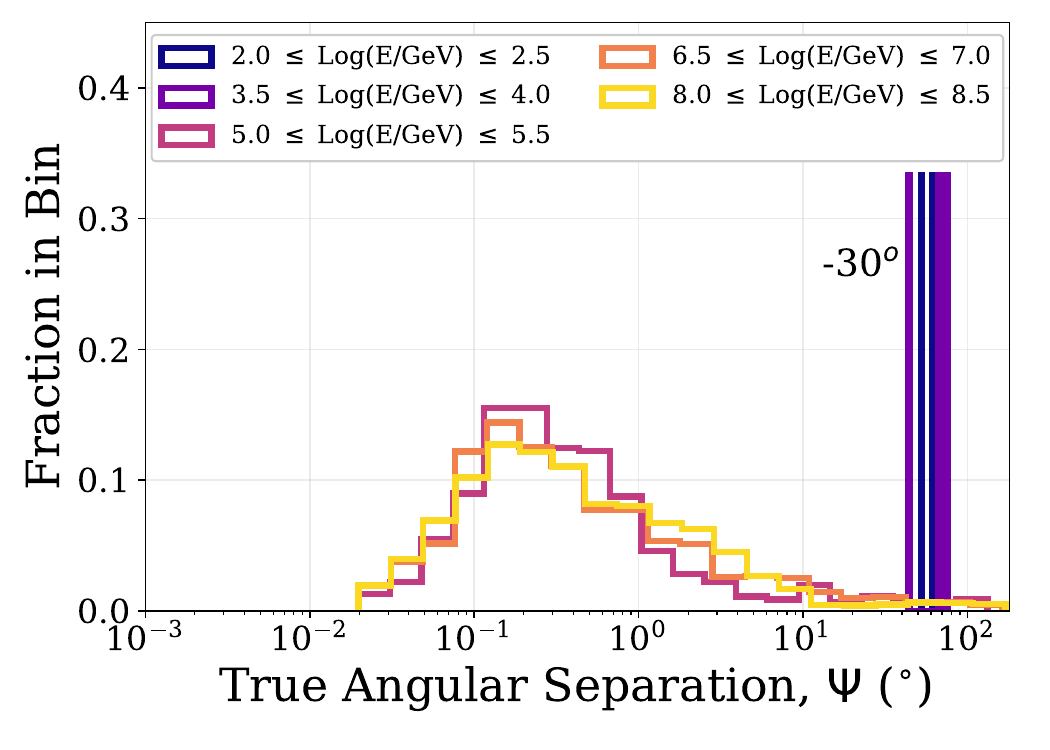}
  \caption{Examples of the binned point spread functions measured in the northern sky ($\delta = 30^{\circ}$), horizon ($\delta = 0^{\circ}$), and southern sky ($\delta = -30^{\circ}$) for the IC86 season. Each colored histogram corresponds to a different true neutrino energy range. For a falling $E^{-2}$ spectrum, most muons are reconstructed less than $1^{\circ}$ from the neutrino origin. Note that in the rightmost plot, very few low-energy events pass selection criteria in the southern sky, resulting in sparsely populated PDFs in this region.}
  \label{fig:irf_psf}
\end{figure}

\begin{figure}
  \centering
  \includegraphics[width=0.32\textwidth]{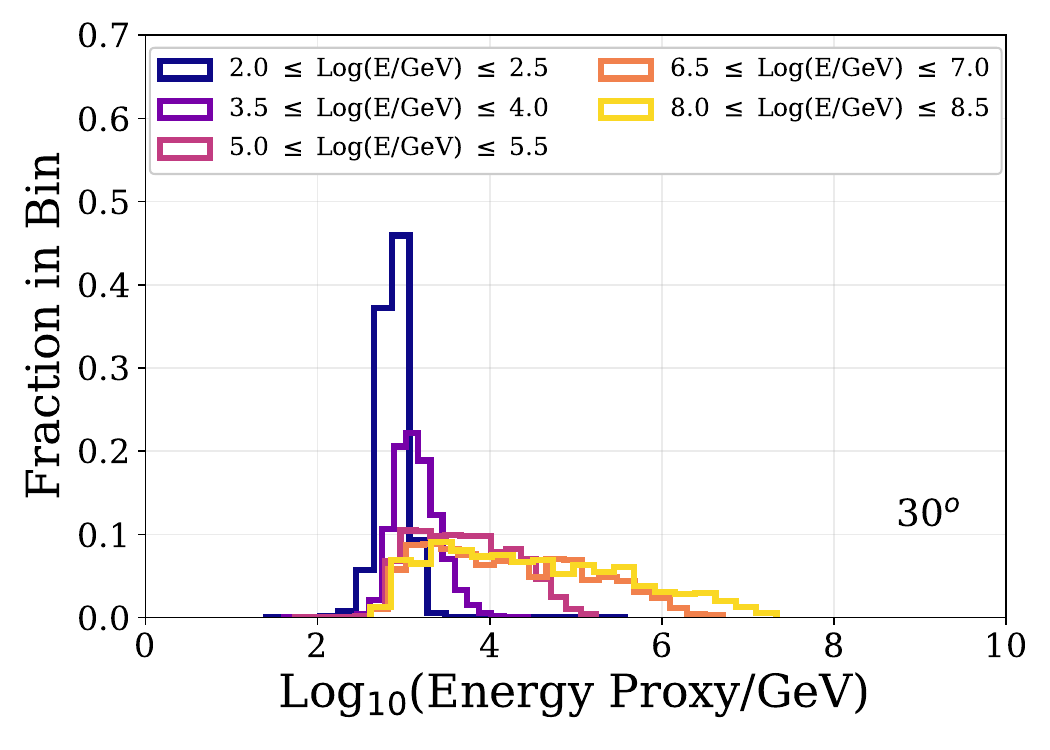}
  \includegraphics[width=0.32\textwidth]{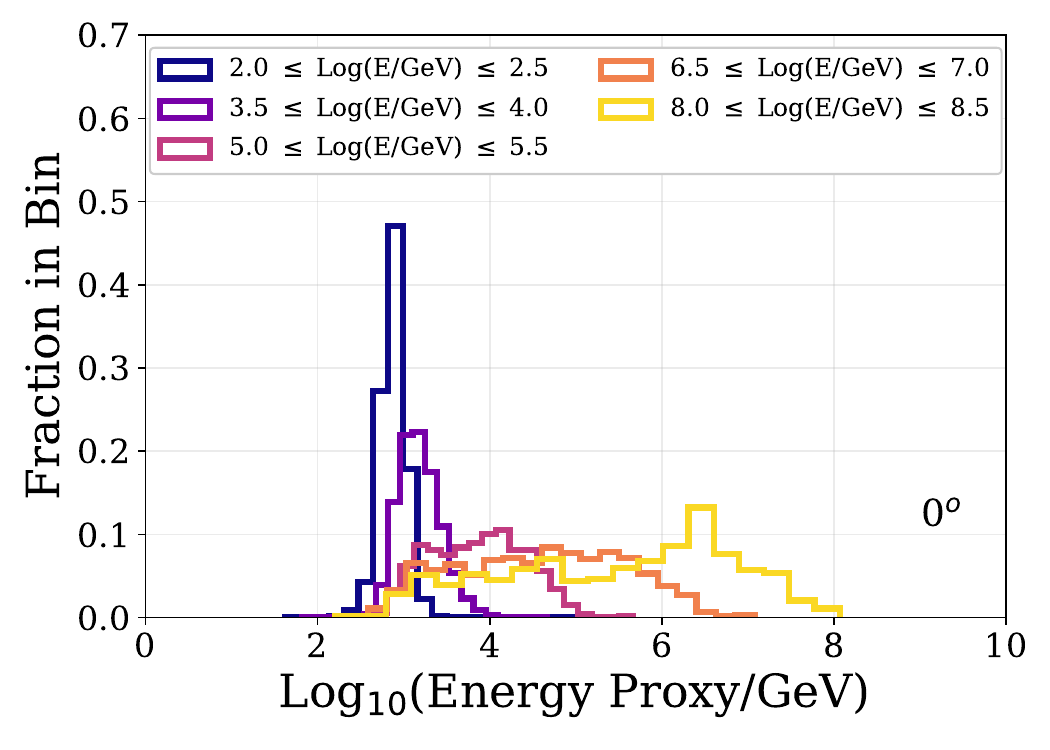}
  \includegraphics[width=0.32\textwidth]{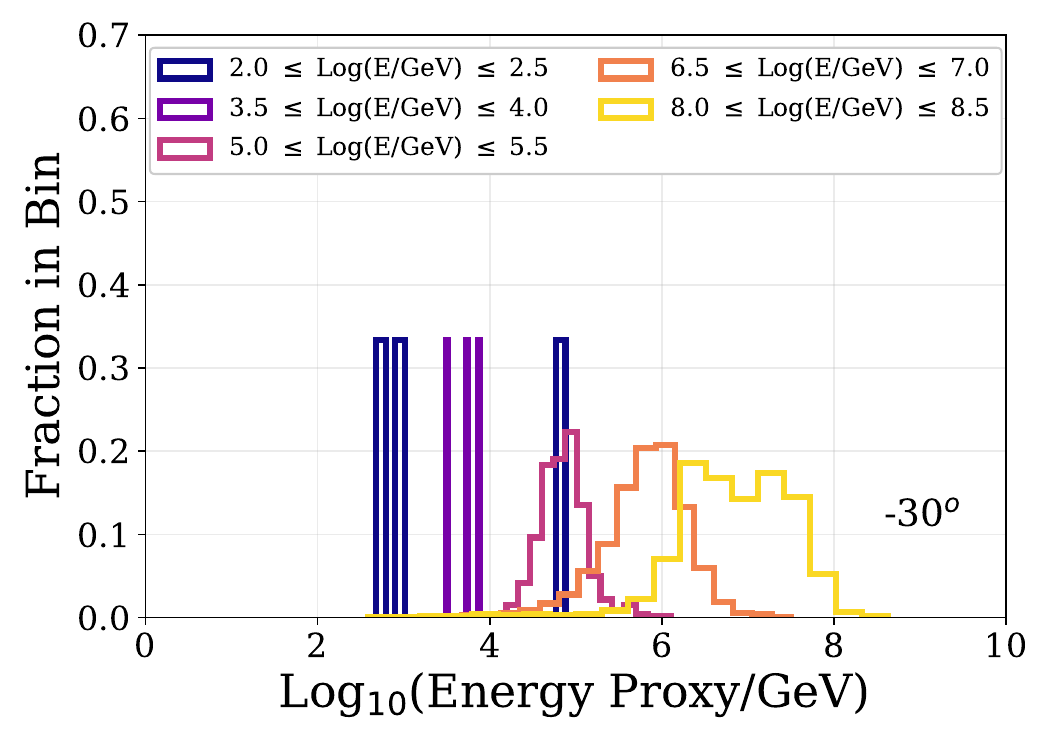}
  \caption{Examples of the binned reconstructed muon energy measured in the northern sky ($\delta = 30^{\circ}$), horizon ($\delta = 0^{\circ}$), and southern sky ($\delta = -30^{\circ}$) for the IC86 season. Each colored histogram corresponds to a different true neutrino energy range. In the southern sky, high energy events reconstruct near the incident neutrino energy. In the northern sky and at the horizon, high energy events may interact far from the detector, producing energy losses which are not visible in IceCube.}
  \label{fig:irf_esmearing}
\end{figure}

\begin{figure}
  \centering
  \includegraphics[width=0.32\textwidth]{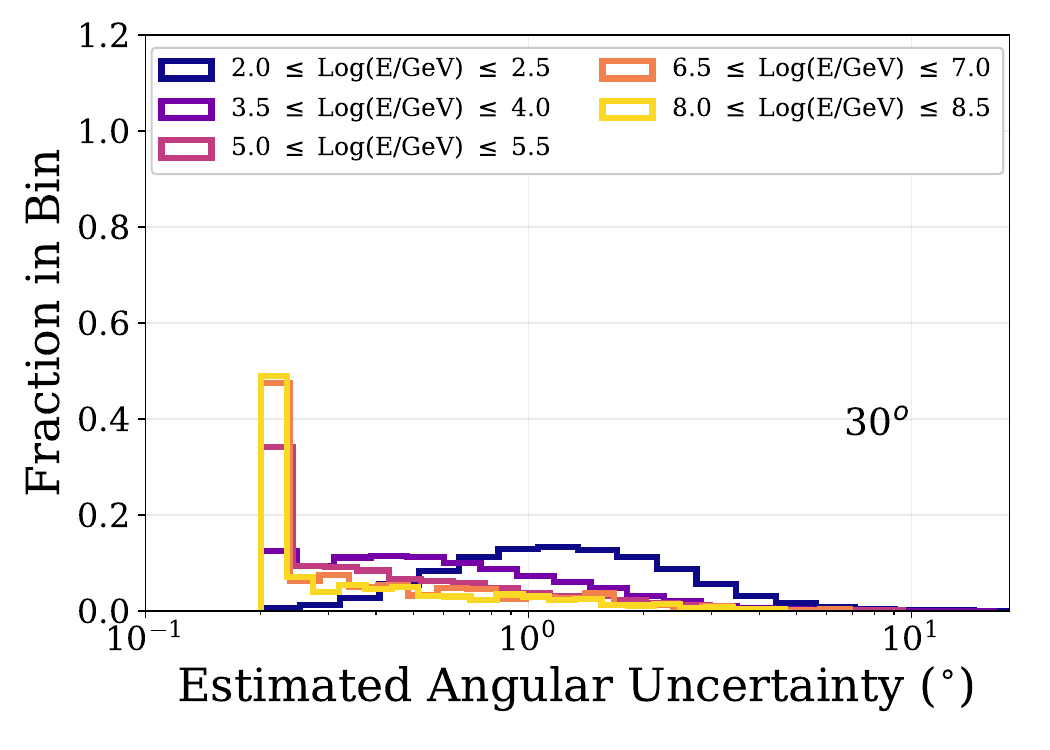}
  \includegraphics[width=0.32\textwidth]{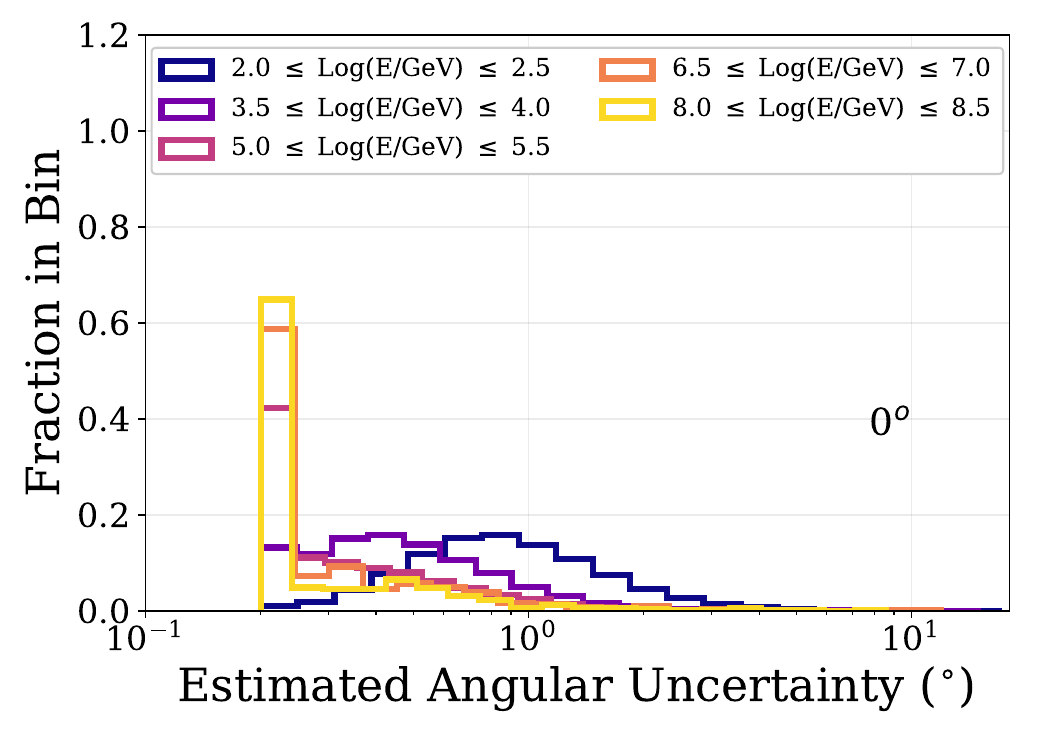}
  \includegraphics[width=0.32\textwidth]{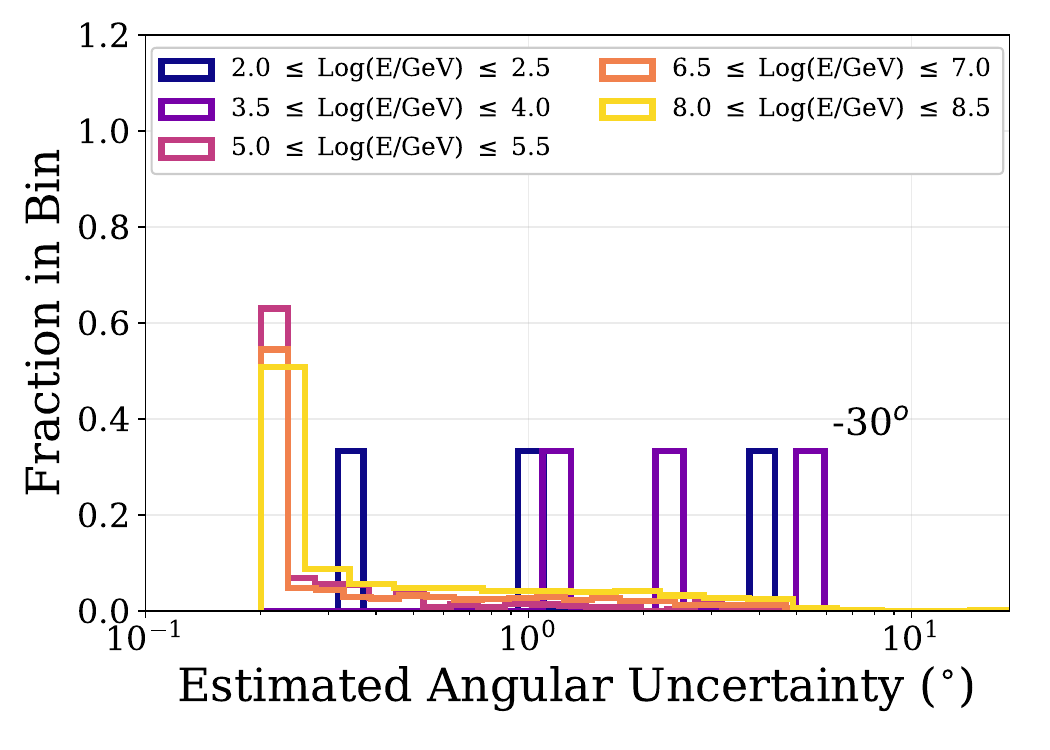}
  \caption{Examples of the binned estimated angular uncertainty on the reconstructed direction measured for the northern sky  ($\delta = 30^{\circ}$), horizon ($\delta = 0^{\circ}$), and southern sky ($\delta = -30^{\circ}$) for the IC86 season. Each colored histogram corresponds to a different true neutrino energy range. The estimated angular uncertainties have been calibrated to give correct coverage on average for an $E^{-2}$ spectrum. A floor of 0.2$^{\circ}$ is included for all events to avoid potential unaccounted-for systematic uncertainties dominating the generic point source likelihood.
}
  \label{fig:irf_angerr}
\end{figure}

Figures ~\ref{fig:irf_psf}, \ref{fig:irf_esmearing}, and \ref{fig:irf_angerr} show distributions of the point spread function, reconstructed muon energy, and angular uncertainty in three different regions of the sky. Note that the event selection attempts to remove low energy events in the southern sky, leading to low statistics and mostly misreconstructed events in this region.

\section{Internal Due Diligence Results}\label{sec:due-diligence}

\subsection{Time-Integrated Searches}
\label{sec:time-int-subsec}
\subsubsection{Analysis Methods}
Checks with the time-integrated point-source analysis use the same maximum-likelihood framework described in~\citep{Aartsen:2013uuv}. The likelihood has the form
\begin{equation}\label{eq:time_int_llh}
\mathcal{L}(\tilde{n}_s, \tilde{\gamma})=\prod_{i=1}^N\left[\frac{n_s}{N}\mathcal{S}_i(\mathbf{x}_S, \mathbf{x}_i, \sigma_i, E_i; \gamma)+\left(1-\frac{n_s}{N}\right)\mathcal{B}_i(\sin \delta_i, E_i)\right].
\end{equation}

In this expression, $\mathcal{L}$ is maximized with respect to $n_s$, the estimated number of signal events, and $\gamma$, the best-fit spectral index at the tested source location. The $\mathcal{S}$ and $\mathcal{B}$ PDFs each contain spatial and energy terms that quantify how likely an event is to be signal or background respectively. The signal term depends on the source coordinates ($\mathbf{x}_S$), the reconstructed event direction ($\mathbf{x}_i$), the event’s angular uncertainty ($\sigma_i$), and the reconstructed energy ($E_i$) assuming a given source spectrum ($E_i; \gamma$). The background term depends on $\sin{\delta_i}$ and $E_i$. For the track dataset used in this analysis, the background is well described by the total event rate within each declination or zenith band. The full likelihood is the product of the per-event terms for all $N$ events in the dataset.

The Test-Statistic (TS) for the time-integrated search is defined as the ratio of the likelihood under the null hypothesis ($n_s=0$) to the likelihood at the best-fit signal parameters:
\begin{equation}\label{eq:TI_TS}
\mathrm{TS}=-2\log\left[\frac{\mathcal{L}(n_s=0)}{\mathcal{L}(\hat{n}_s, \hat{\gamma})}\right].
\end{equation}
Here, $\hat{n}_s$ and $\hat{\gamma}$ are the best-fit values obtained from the likelihood maximization.

For all-sky point-source searches, the pre-trial, or local, $p$-value is determined by comparing the measured TS to a background distribution produced through many pseudo-experiments in which the real data is scrambled in right ascension. Each scrambled dataset is processed with the same unbinned maximum-likelihood analysis. These background distributions are generated separately for each declination band in 1$^\circ$ spacings from $-80^\circ$ to $+80^\circ$. The pre-trial $p$-value is the fraction of background trials that produce a TS greater than the observed value.

For the all-sky, time-integrated search, two post-trial $p$-values are independently calculated, one for each hemisphere. This is done by creating many background skymap realizations, identifying the most significant location in each hemisphere for every realization, and forming two distributions of these ``hottest-spot" $p$-values. The post-trial $p$-value for the measured skymap is the fraction of background hottest-spot values in the relevant hemisphere that are larger than the measured one.

For the catalog search, only the most significant source receives a trial corrected significance to account for the look-elsewhere effect. The correction is applied using the Šidák correction method~\cite{sidak}.

\subsubsection{All-Sky Point-Source Search Results}
The results of the time-integrated all-sky point-source search are from the analysis described in~\citep{IceCube:2025arXiv}. The likelihood in Eq.~\ref{eq:time_int_llh} is evaluated across a full-sky grid of pixels, and the corresponding TS from Eq.~\ref{eq:TI_TS} is computed at each grid point. The grid is produced using the \verb|Healpy|~\citep{healpy} Python~\citep{python} implementation of the \verb|HEALPix|\footnote{\url{http://healpix.sourceforge.net}}
 framework~\citep{healpix} with $N_{side}=128$, which gives a pixel area of roughly 0.21~deg$^2$. Regions within 10 degrees of the celestial poles are excluded because the statistics there are poor.

\begin{figure}
  \centering
  \includegraphics[width=.8\textwidth]{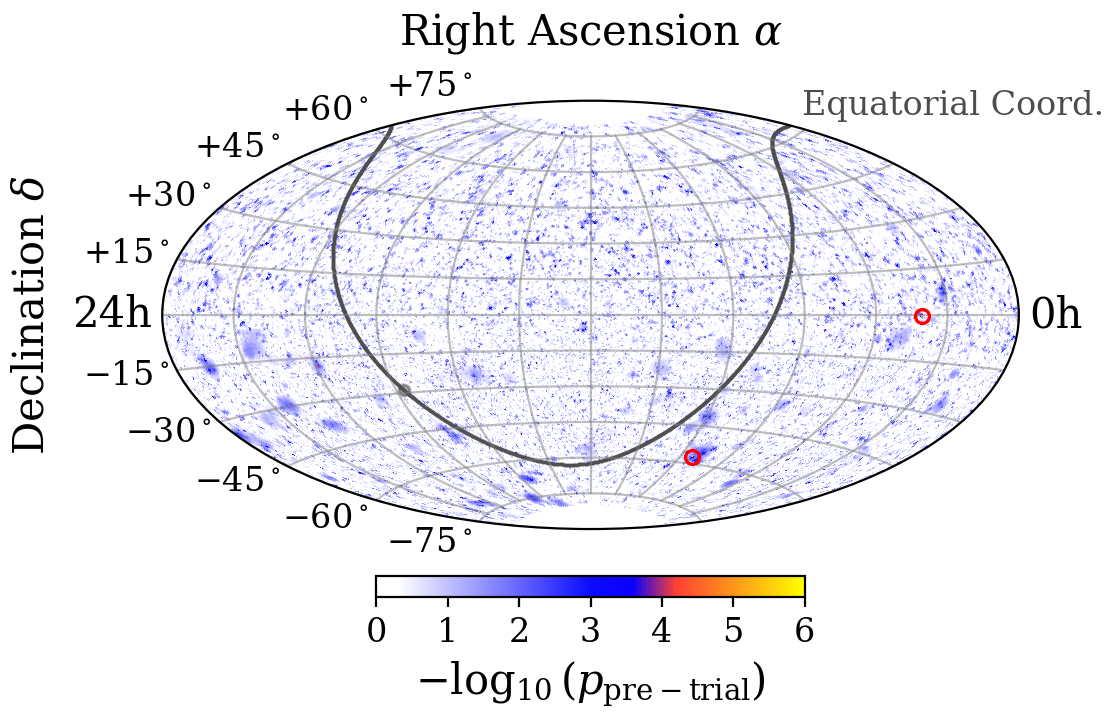}
  \caption{\textbf{Time-Integrated All-Sky Map.} Pre-trial $p$-value map using 14 years of tracks in equatorial coordinates (J2000). The north/south boundary is at declination $-5$ degrees, separating sky regions where different processing strategies apply, as described in section \ref{sec:evt-selection}. This map includes 4 additional years of data compared to the previous IceCube track-based point-source search~\citep{Aartsen:2019fau}. The gray line marks the Galactic plane, and the gray dot marks the Galactic Center. The hottest northern and southern spots are circled in red and described in Table~\ref{tab:generic_tint_skymap}.}
  \label{fig:TI_all_sky_scan}
\end{figure}

A pre-trial $p$-value is calculated at the center of every pixel. In each hemisphere, the pixel with the smallest $p$-value is taken as the hottest spot. The post-trial $p$-value is calculated for the hottest northern and southern spots. Neither the northern nor southern hottest spot reaches the $3\sigma$ evidence threshold. The fitted values of $\hat{n}_s$, $\hat{\gamma}$, and the equatorial coordinates of the hottest spots are listed in Table~\ref{tab:generic_tint_skymap}. The complete pre-trial significance map for tracks and cascades is shown in Fig.~\ref{fig:TI_all_sky_scan}.

The hottest spot in the northern sky is located at $\delta = -0.30^\circ$ and $\mathrm{R.A.}= 40.8^\circ$. After trial correction, the post-trial $p$-value is 0.22. This pixel is approximately $0.3^\circ$ from the galaxy NGC 1068. The fitted parameters are $\hat{n}_s = 69$ and $\hat{\gamma} = 3.3$.

\begin{figure}
  \centering
  \includegraphics[width=.8\textwidth]{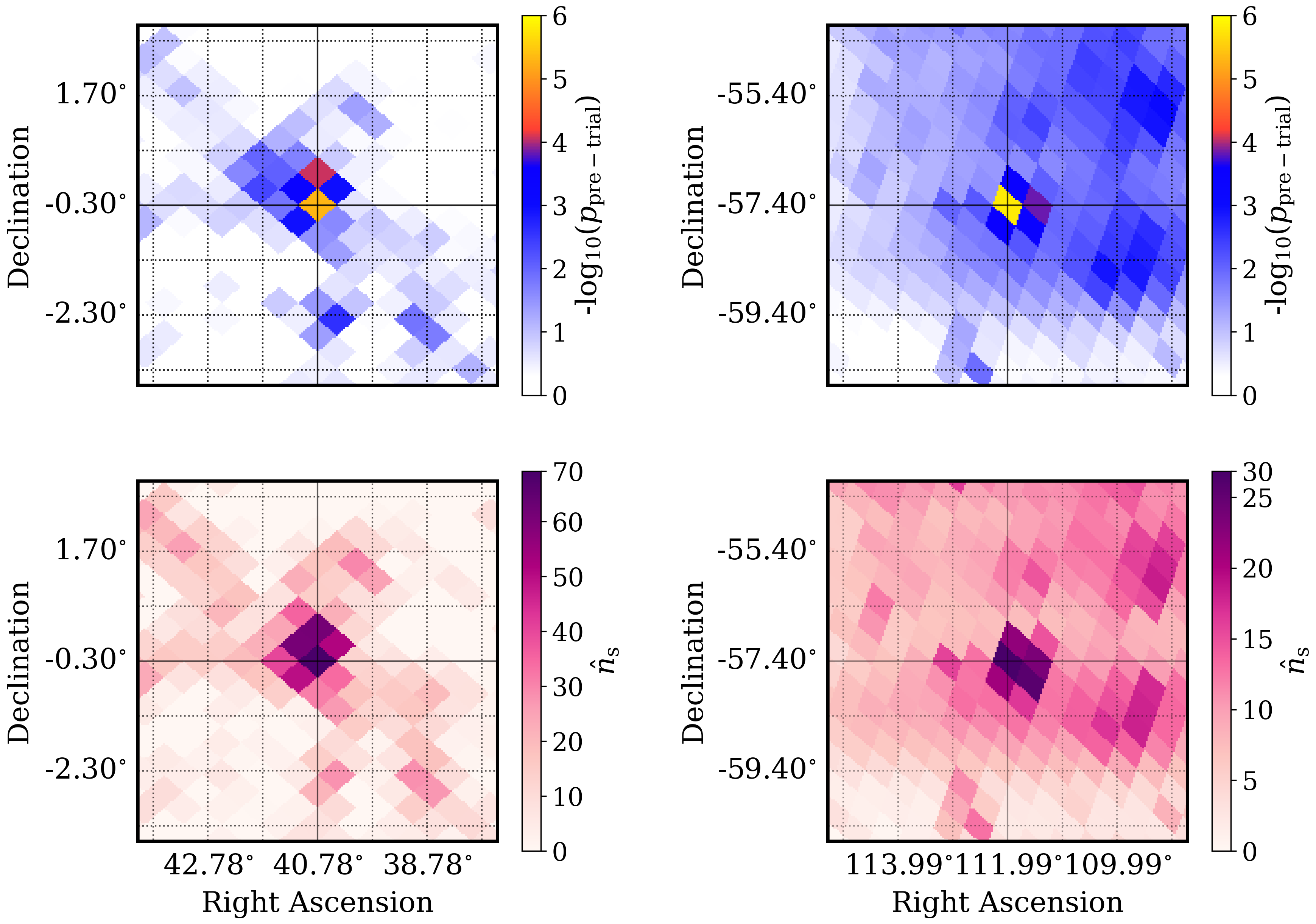}
  \caption{\textbf{Northern and Southern Hottest Spots.} Pre-trial $p$-value (top row) and $n_s$ (bottom row) maps in equatorial coordinates of the area around the most significant point in the northern sky (left column) and southern sky (right column). }
  \label{fig:both_hotspot_zooms}
\end{figure}

In the southern sky, the hottest pixel is found at $\delta = -57.4^\circ$ and $\mathrm{R.A.} = 112^\circ$, with a post-trial $p$-value of 0.05. The fitted parameters for this location are $\hat{n}_s = 27$ and $\hat{\gamma} = 1.9$. Figure~\ref{fig:both_hotspot_zooms} shows $3^\circ \times 3^\circ$ maps of $-\log{10}(p_{\mathrm{local}})$ and the fitted $n_s$ values around each hottest spot. Both hottest spots are consistent with the background under this search procedure.

\begin{table*}[htbp]
    \centering
    \setlength{\tabcolsep}{25pt}
    \begin{tabular}{c  c  c  c  c  c} 
     \toprule
       & Post-trial $p$-value & $n_s$ & $\gamma$ & R.A. [deg] & Decl. [deg] \\ [0.5ex] 
     \midrule
     \midrule
     North & 0.22 & 69 & 3.3 & 40.8 & -0.30\\  
     South & 0.05 & 27 & 1.9 & 112  & -57.4\\ 
     \bottomrule
    \end{tabular}
    \caption{\textbf{All-sky Point-Source Search Hottest Northern and Southern Spots.} Summary of global $p$-value, number of signal events ($n_s$),  power-law spectral index ($\gamma$), and location (right ascension and declination) of the most significant point in the northern and southern sky. The global $p$-value is calculated by correcting for testing locations across the source's corresponding part of the sky (either northern or southern).}
    \label{tab:generic_tint_skymap}
\end{table*}

\subsubsection{Time-Integrated Source Catalog Results}\label{sec:time-int-catalog}
A targeted search is performed on a set of previously identified $\gamma$-ray emitters using the tracks dataset. The catalog contains 110 sources, identical to the list used in \citet{Aartsen:2019fau}. A complete table of these sources and their corresponding $p$-values is provided in~\citep{IceCube:2025arXiv}. Table~\ref{tab:source_list} summarizes the pre-trial $p$-values, equatorial coordinates, and best-fit values of $\gamma$ and $n_s$ for the four most significant sources.

The highest-significance source in the catalog is NGC 1068. In this analysis, NGC 1068 has a pre-trial $p$-value of $2.07 \times 10^{-6}$ (4.6$\sigma$), obtained from a set of 30 million scrambled background trials. After correcting for the 110 source trials in the catalog, the global $p$-value is $2.28 \times 10^{-4}$ (3.5$\sigma$). The fitted values at this position are $\hat{n}_s = 71.1$ and $\hat{\gamma} = 3.14$; consistent with those obtained for the northern hottest spot in the all-sky scan.

The results from~\citet{IceCube:2025arXiv} presented here are consistent with previous analyses indicating that NGC 1068 is a promising neutrino emitter~\citep{Aartsen:2019fau, IceCube:2022Science}. The differences in significance relative to \citet{Aartsen:2019fau} arise from the additional years of track data used in this work, while differences relative to \citet{IceCube:2022Science} are due to changes in track reconstruction methods.
\begin{table}[!htb]
    \centering
    \setlength{\tabcolsep}{13pt}
    \begin{tabular}{lcccccccc}
    \toprule
     & Source Name & Class & RA [deg] & Dec [deg] & $\hat{n}_s$ & $\hat{\gamma}$ & Pre-trial $p$-value \\
    \midrule
    \midrule
    1 & NGC 1068 & SBG & 40.67 & -0.01 & 71.10 & 3.14 & $2.07 \times 10^{-6}$ \\
    2 & PKS 1424+240 & BLL & 216.76 & 23.8 & 65.52 & 3.44 & $1.36 \times 10^{-4}$ \\
    3 & TXS 0506+056 & BLL & 77.35 & 5.7 & 9.42 & 1.89 & $1.61 \times 10^{-3}$ \\
    4 & GB6 J1542+6129 & BLL & 235.75 & 61.5 & 41.31 & 3.33 & $3.18 \times 10^{-3}$ \\
    \bottomrule
    \end{tabular}
    \caption{\textbf{Most Significant Time-Integrated Source-List Objects.} Summary of location, number of signal events, $n_s$, power-law spectral index, $\gamma$ and pre-trial $p$-value of the four most significant sources in the catalog.}
    \label{tab:source_list}
\end{table}

\subsection{Time-Dependent Searches}
\subsubsection{Analysis Methods}
In this section, we carry out the time-dependent counterpart to the analysis in \ref{sec:time-int-subsec}. We work with a likelihood that has the same structure as Eq. \eqref{eq:time_int_llh}. 
This time, both $\mathcal{S}$ and $\mathcal{B}$ include an additional temporal PDF. For the signal, we assume a flat time PDF in 
$[T_0-\Delta T/2, T_0+\Delta T/2]$, where $T_0$ and $\Delta T$ are the center and width of the hypothetical flare, respectively. On the other hand, the background time PDF is flat throughout the livetime of the sample. The likelihood is maximized over $n_s$, $\gamma$, $T_0$ and $\Delta T$ to compute the TS.

\begin{figure}
    \begin{center}
\includegraphics[width=0.9\textwidth]{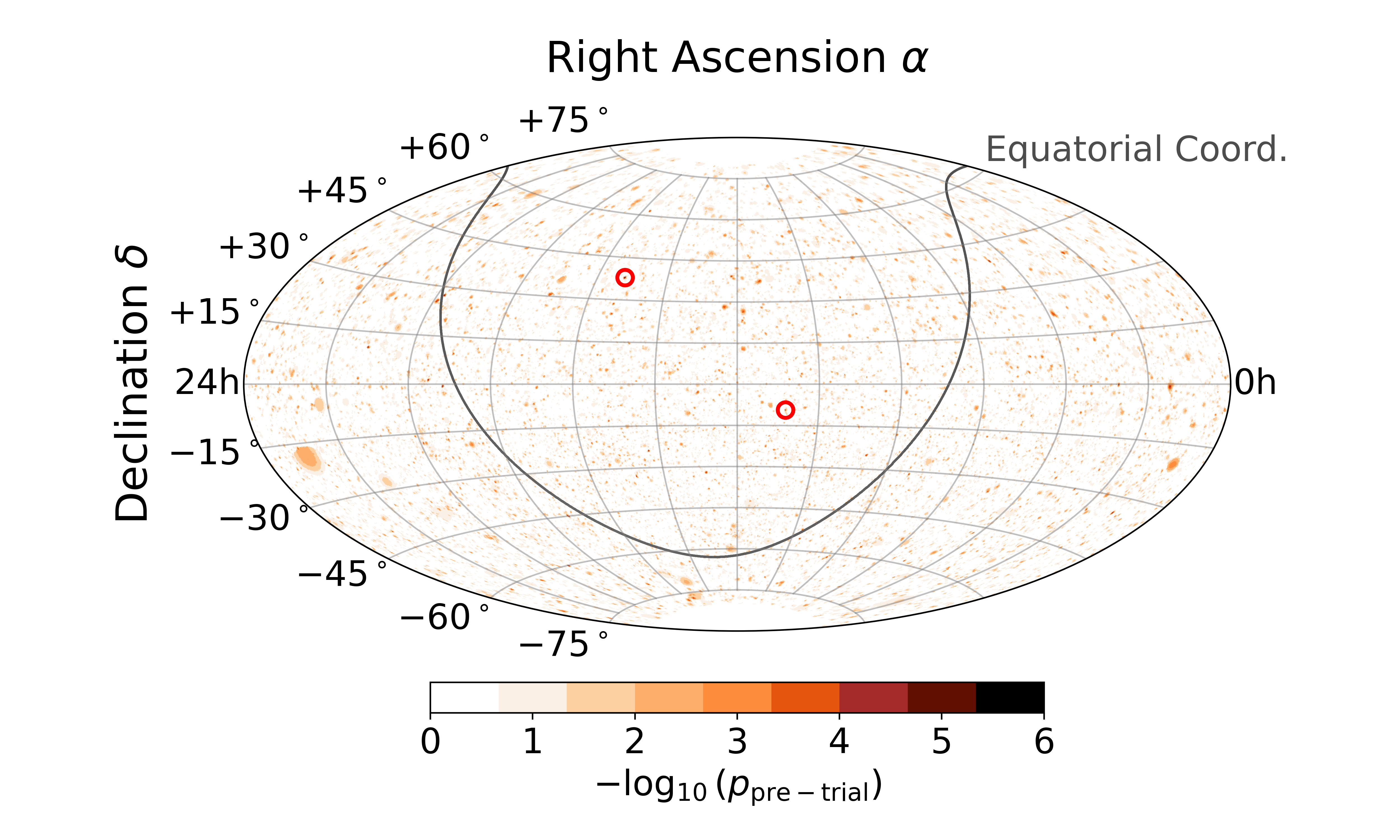}
    \caption{\textbf{Time-Dependent All-Sky Map.} The time-dependent counterpart to Figure~\ref{fig:TI_all_sky_scan}. Northern and southern sky hotspots are marked as red circles.}
    \label{fig:timedep_skymap}
    \end{center}
\end{figure}

\subsubsection{Time-Dependent Skymap Results}
We divide the skymap into pixels with the parameter N$_{\mathrm{side}}$=256, yielding a mean size of $\approx 0.23^\circ$. 
Similar to the time-dependent all-sky scan, we report the hotspots in the northern and southern sky, which we summarize in Table~\ref{tab:generic_tdep_skymap}. Neither hotspot is statistically  significant after trial corrections. The northern hotspot is not near any tested source. The third most significant hotspot is spatially close to the location of NGC 1068, however, with $\mathrm{R.A.}=40.8^\circ$ and $\delta=-0.15^\circ$, and pre-trial $p$-value $p=1.47\times 10^{-5}$.

\begin{table}[htb]
    \centering
    \setlength{\tabcolsep}{19pt}
    \begin{tabular}{cccccccc}
    \toprule
      & R.A.  & Decl. & $\hat{n}_s$ & $\hat{\gamma}$ & $\hat{T}_0$  & $\Delta \hat{T}$ & Pre-trial $p$-value \\
        & [deg] & [deg] & & & [MJD] & [days] & \\
    \midrule
    \midrule
    North & 228.3 & 37.9 & 60 & 2.95 & 59199 & 911 & $3.72\times10^{-6}$ \\
    South & 162.2 & -9.3 & 13 & 3.20 & 56584 & 393 & $1.60\times10^{-5}$ \\
    \bottomrule
    \end{tabular}

    \caption{\textbf{Northern and Southern Sky Hotspots from the Time-Dependent All-Sky Scan.} We summarize the location of the hotspots, their associated pre-trial $p$-value and fit parameters.}
    \label{tab:generic_tdep_skymap}
\end{table}

\subsubsection{Time-Dependent Catalog Results}\label{sec:tim-dep-catalog}

Using the same source catalog in Section \ref{sec:time-int-catalog}, we compute the pre-trial $p$-values in a time-dependent search. We summarize our results in Table~\ref{tab:tim-dep-catalog}. Our most significant source is also NGC 1068, with  a very long flare of $\Delta \hat T $ = 2956 days. The TXS 0506+056 neutrino flare in this search is consistent with the results from \citet{IceCube:2018Science_flare}.

\begin{table}
    \setlength{\tabcolsep}{9pt}
    \centering
    \begin{tabular}{ccccccccc}
    \toprule
    Source Name  & R.A.  & Decl. & $\hat{n}_s$ & $\hat{\gamma}$ & $\hat{T}_0$  & $\Delta \hat{T}$ & Pre-trial $p$-value & Post-trial $p$-value\\
        & [deg] & [deg] & & & [MJD] & [days] & \\
    \midrule
    \midrule
     NGC 1068 & 40.67 & -0.01 & 65.29 & 3.02 & 57410 & 2956 & $8.95 \times 10^{-6}$ & $9.84\times 10^{-4}$\\
     WComae & 185.38 & 23.24 & 12.94 & 2.76 & 55692 & 23 & $1.67 \times 10^{-4}$ & $1.84\times 10^{-2}$\\
     GB6 J1542+6129  & 235.75 & 61.5 & 29.4 & 2.67 & 57761 & 411 & $2.76 \times 10^{-4}$ & $3.03\times 10^{-2}$ \\
    PKS1424+240	& 216.76 & 23.8 & 70.86 &3.27 & 57559 &	3645 & $1.44\times 10^{-3}$ & 0.147\\
    TXS 0506+056 & 77.35 & 5.7 & 12.56 & 2.26 & 57020 & 185 & $4.18 \times 10^{-3}$ & 0.369\\
    \bottomrule
    \end{tabular}
    \caption{\textbf{Most Significant Sources from the Time-Dependent Catalog Search.} We summarize their locations, pre-trial and post-trial $p$-values and fit parameters.}
    \label{tab:tim-dep-catalog}
\end{table}

\subsection{Time-Integrated Galactic Plane Template Search}
\subsubsection{Diffuse Galactic Plane Emission Hypotheses}
The Galactic plane analysis uses the same unbinned maximum-likelihood framework described in Eq.~\ref{eq:time_int_llh}, but with two modifications. First, the spatial PDF is now constructed by taking a specified Galactic plane neutrino emission model, normalizing it to produce spatial PDF and then convolving it with both the IceCube detector acceptance and each event's estimated angular uncertainty. The latter convolution is performed via Gaussian smearing with widths corresponding to each event's estimated angular resolution. Second, the fit includes only one free parameter, the number of signal events $n_s$. The spectrum is fixed. Therefore, the test statistic reduces to

\begin{equation}\label{GP_TS}
    \mathrm{TS} = -2\log\left[\frac{\mathcal{L}(n_s=0)}{\mathcal{L}(\hat{n}_s)}\right].
\end{equation}
The templates evaluated in this work are the same as those tested in~\citep{IceCube:2023Science}: \textit{Fermi}-LAT $\mathbf{\pi^0}$, KRA$\mathbf{^5_\gamma}$, and KRA$\mathbf{^{50}_\gamma}$. 

The \textit{Fermi}-LAT $\pi^0$ template represents the $\pi^0$ component of the diffuse gamma-ray sky measured by \textit{Fermi}-LAT~\citep{Fermi-LAT}. In hadronic interactions between cosmic rays and the interstellar medium, $\pi^0$ mesons decay to gamma rays, while $\pi^\pm$ produce neutrinos. Because these processes occur together, the $\pi^0$ map provides a proxy for the expected diffuse neutrino emission. This model assumes uniform cosmic-ray propagation throughout the Galactic disk. The spatial distribution follows the gas density profile of the Galactic plane, and the associated neutrino spectrum is approximated as a power law with spectral index $\gamma = 2.7$.

The KRA${^5_\gamma}$ and KRA${^{50}_\gamma}$ templates follow the model developed by~\citep{Gaggero_2015}, which was introduced to address the underestimation of high-energy gamma rays in the inner Galaxy observed by both the Milagro experiment~\citep{7yrPSTracks,Gaggero_2015} and more recently by \textit{Fermi}-LAT. In this scenario, cosmic-ray transport varies with Galactic radius, producing harder spectra toward the Galactic center.
The KRA${^5_\gamma}$ and KRA${^{50}_\gamma}$ templates assume a cutoff in the cosmic ray spectrum at either 5~PeV or at 50~PeV, respectively. A cutoff with 50~PeV is considered optimistic, while the 5~PeV is more similar to current models, but both were tested in order to provide limits for the available models. The KRA$_\gamma$ models used in this analysis were made available by the authors~\citep{Gaggero_2015} and can also be accessed publicly~\citep{gaggero_public}. These models have been re-binned across the full sky and energy domain and used as both spatial and spectral templates, as was done in previous IceCube analyses~\citep{Aartsen_2019,7yrPSTracks, IceCube:2023Science}.

\subsubsection{Galactic Plane Emission Results}
Here we report the pre-trial $p$-values for all three templates tested with 15 years of tracks.  Although the three Galactic Plane emission hypotheses exhibit some correlation, a conservative trials factor of 3 is applied to only the most significant template to calculate the post-trial $p$-value.

The KRA$_\gamma^{5}$ template yielded the highest significance in these tests, with a pre-trial significance of 1.74$\sigma$.  Therefore, the post-trial significance of the KRA$_\gamma^{5}$ template is 1.16$\sigma$. A summary of results for all tests conducted in this analysis are presented in Tab.~\ref{tab:GP}. The tracks Galactic plane tests do not provide significant evidence for any of the three templates tested.

\begin{table}[!htb]
    \centering
    \setlength{\tabcolsep}{20pt}
    \renewcommand{\arraystretch}{1.4}
    \begin{tabular}{cccccc}
    \toprule
    Template & $\hat{n}_s$ & Sensitivity Flux & Best-fit Flux & Pre-trial $p$-value (significance) \\
    \midrule
    \midrule
    \textit{Fermi}-LAT $\pi^0$ & 727 & 10.2 & 12.0$^{+8.4}_{-8.4}$ & 0.07 (1.45$\sigma$) \\
    KRA$_\gamma^{5}$ & 397 & 0.51$\times$MF & 0.66$^{+0.40}_{-0.38}\times$MF & 0.04 (1.74$\sigma$) \\
    KRA$_\gamma^{50}$ & 283 & 0.36$\times$MF & 0.42$^{+0.22}_{-0.34}\times$MF & 0.05 (1.64$\sigma$)\\
    \bottomrule
    \end{tabular}
    \caption{\textbf{$\mathbf{\hat{n}_s}$, Flux or Relative Model Normalization, and pre-trial $p$-value with gaussian-equivalent significance in parentheses.} Summary of the template best-fit number of signal events ($\hat{n}_s$), template best-fit flux/model norm, pre-trial $p$-value, and pre-trial significance of the three templates tested with 15yr tracks sample. For the \textit{Fermi}-LAT $\pi^0$ template, fluxes are shown as E$^2\frac{dN}{dE}$ at 100~TeV in units of $10^{-11}$\;TeV cm$^{-2}$ s$^{-1}$. For the KRA$_\gamma$ templates, the flux is shown in units of predicted model flux (MF).}
    \label{tab:GP}
\end{table}

\section{Benchmark Results Comparison Using Public Tools}\label{sec:benchmarkresults}
To validate the detector response information included in IceTracks-DR2, we used the publicly available SkyLLH\footnote{\url{https://github.com/icecube/skyllh}} package to reproduce key benchmark results based on the provided event sample, response matrices, and effective areas. SkyLLH is an open-source, Python-based likelihood analysis framework~\citep{Wolf:2019ICRC,Kontrimas:2021icrc} developed within the IceCube collaboration to perform frequentist statistical data analyses for point-like neutrino source searches. It has been previously extended to support the analysis of IceTracks-DR1~\citep{Bellenghi:20230u} through a public data interface which enables users to construct signal and background probability density functions (PDFs) directly from the released IRFs~\citep{Bellenghi:20230u}.
The only component of the likelihood that differs between the public implementation and internal IceCube tools is the signal energy PDF. In internal analyses, this PDF is parameterized as $P(\tilde E_\mu|\sin(\tilde\delta_{\mu}),\gamma)$, where $\tilde E_\mu$ and $\tilde\delta_{\mu}$ are the reconstructed muon energy and declination.
Using the public information, the same PDF can instead be constructed from the response matrix as conditionally dependent on the parent neutrino declination, $\delta_\nu$. Under the assumption of an unbroken power law energy spectrum, it takes the form

\begin{equation}\label{eq:energy_pdf}
    P_{\rm{E}}(\tilde E_\mu|\delta_{\mathrm{src}},\gamma) \equiv P(\tilde E_{\mu}|\delta_{\nu},\gamma) = \int_{E_{\mathrm{min}}}^{E_{\mathrm{max}}} dE_{\nu}\,P(\tilde E_{\mu}|E_{\nu},\delta_{\nu})\,P(E_{\nu}|\gamma)\,P(E_{\nu}|\delta_{\nu}),
\end{equation}

The integrand of \autoref{eq:energy_pdf} is the product of three probabilities: $P(\tilde E_{\mu}|E_{\nu},\delta_{\nu})$ is calculated from the detector response matrix; $P(E_{\nu}|\gamma)$ is known for a power-law energy spectrum with given spectral index $\gamma$; $P(E_{\nu}|\delta_{\nu})$ is calculated from the provided tabulated effective areas. The integration limits $E_{\rm{min}}$ and $E_{\rm{max}}$ are set in order to cover the parent neutrino energy range of $10^2 - 10^9$~GeV, as provided in the IRFs.

Although this difference in parameterization may introduce small discrepancies, the primary limitation in IceTracks-DR1 was the coarse declination binning of the response matrices~\citep{IceCube:2021arXiv}. To rectify this issue, IceTracks-DR2 uses the same finer declination binning used to produce PDFs for internal analyses, substantially reducing previous mismatches.
Additional improvements include several minor fixes in SkyLLH (documented in the release notes) and the cleanup of spurious contributions to the binned effective areas. The latter consists of removal of low energy signal events that enter the selection solely due to contributions from coincident air showers, leading to erroneous estimates of the effective areas in these regions.

\begin{figure}
  \centering
  \includegraphics[width=.8\textwidth]{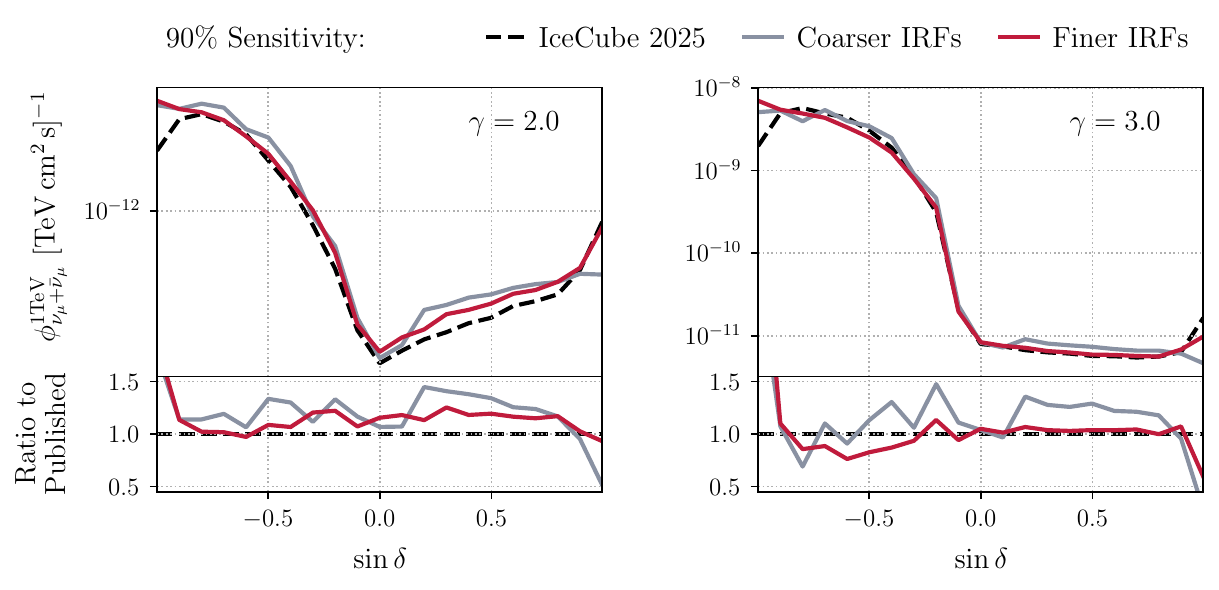}  
  \includegraphics[width=.8\textwidth]{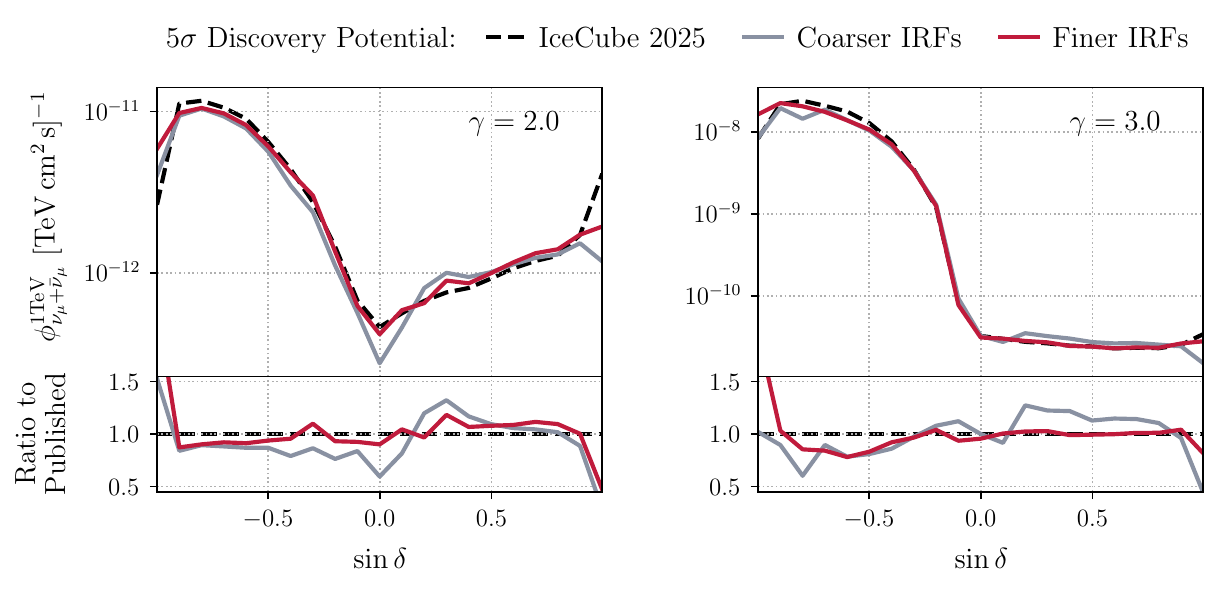}
  \caption{Comparison of the average astrophysical neutrino flux required to reach the 90\% sensitivity (top) and a  5$\sigma$ discovery prior to any penalty factors calculation (bottom) in 14 years of IceCube observations, assuming unbroken power-law spectra with spectral index $\gamma = 2.0$ (left) and $\gamma = 3.0$ (right), shown as a function of source declination. The IceCube sensitivity (black dashed) from \citet{IceCube:2025arXiv} is compared to two benchmark estimates obtained using publicly available tools. The grey curve shows the sensitivity derived using the IRF matrices coarsely binned, as in IceTracks-DR1~\citep{IceCube:2021arXiv}. The red curve shows the result obtained using the more finely binned IRF matrices with misreconstructed events manually removed from the effective area, as described in \autoref{sec:detector-response}.}
  \label{fig:sensitivity-dp}
\end{figure}

The effect of all aforementioned improvements is illustrated by Figure~\ref{fig:sensitivity-dp}. This displays the median expected 90\% flux upper limit in case of a non-observation and the median flux required for a 5$\sigma$ discovery, respectively, as a function of the source declination and for two different assumed spectral indices for the simple power law signal spectrum. These fluxes have been calculated for a signal integrated over the 14-year data sample using both the previous, sparse response matrix declination binning and the finer binning provided in IceTracks-DR2. Both are then compared to the same quantities obtained using per-event detector responses and internal IceCube tools~\citep{IceCube:2025arXiv}. For both spectral assumptions, IceCube's internally produced 90\% sensitivity and 5$\sigma$ discovery potential are better reproduced using the improved IRFs in
IceTracks-DR2.

Apart from making sure that average quantities can be reproduced, we also assess the reproducibility of known excesses. Using SkyLLH, we fit the time-integrated neutrino signal from the Seyfert galaxy NGC\,1068~\citep{IceCube:2022Science,IceCube:2025ApJ,IceCube:2025arXivSeyferts} and from the other 3 top sources from the catalog search reported in section~\ref{sec:time-int-catalog} and in the related publication~\citep{IceCube:2025arXiv}. We maximize the likelihood ratio with respect to the single power law flux parameters $n_{\mathrm{s}}$ and $\gamma$, and estimate the significance of the observed excesses. All results, summarized in Table~\ref{tab:reproducibility_time_int}, are compatible with prior IceCube publications within uncertainties due to differences in the IRFs binning and the signal energy PDF definition explained earlier in this section.

\begin{table}[htbp]
    \begin{tabular}{l|cc|cc|cc}
        \toprule
        \multirow{2}{*}{Source}
        & \multicolumn{2}{c|}{$\hat{n}_s$}
        & \multicolumn{2}{c|}{$\hat\gamma$}
        & \multicolumn{2}{c}{Pre-trial $p$-value (significance)} \\
        \cmidrule(lr){2-3} \cmidrule(lr){4-5} \cmidrule(lr){6-7}
        & DR2 & Published
        & DR2 & Published
        & DR2 & Published \\
        \midrule
        \midrule
        \ngc & 79.5 & 71.1 & 3.2 & 3.1 & $5.2\times 10^{-7}$ ($4.9\sigma$) & $2.1\times 10^{-6}$ ($4.6\sigma$) \\
        PKS\,1424+240 & 75.2  & 65.5  & 3.6 & 3.4 & $3.3\times10^{-5}$ ($4.0\sigma$) & $1.4\times 10^{-4}$ ($3.6\sigma$) \\
        \txs & 9.0  & 9.4  & 2.0 & 1.9 & $6.1\times10^{-3}$ ($2.5\sigma$) & $1.6\times 10^{-3}$ ($2.9\sigma$) \\
        GB6\,J1542+6129 & 48.1  & 41.3  & 3.6 & 3.3 & $1.1\times10^{-3}$ ($3.1\sigma$) & $3.0\times10^{-3}$ ($2.7\sigma$) \\
        \bottomrule
    \end{tabular}
    \caption{\textbf{Comparison of Best-Fit Parameters.} A comparison of time-integrated best-fit parameters and significances between IceTracks-DR2 and results published by the IceCube collaboration~\citep{IceCube:2025arXiv}. For the 4 most significant candidate sources resulting from the IceCube catalog search (see section~\ref{sec:time-int-catalog}), we list the best-fit flux parameters ($\hat{n}_{\mathrm{s}}$, $\hat\gamma$) and $p$-values (with gaussian-equivalent significance in parentheses). The quoted $p$-values do not include any penalty factor from the look-elsewhere effect, which is not relevant for assessing the reproducibility of the published results.}
    \label{tab:reproducibility_time_int}
\end{table}

\begin{figure}
  \centering
  \includegraphics[width=.4\textwidth]{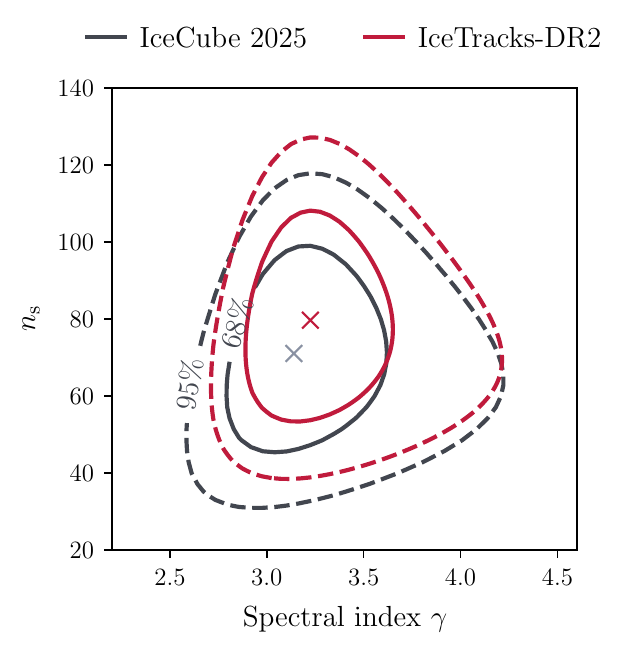}  
  \caption{\textbf{Comparison of NGC 1068 Likelihood Scans.} Comparison of the likelihood scan around the best-fit flux parameters ($\hat{n}_{\mathrm{s}}$, $\hat\gamma$) from the direction of NGC\,1068. The best-fit point (cross) and the 1$\sigma$ (solid line) and 2$\sigma$ (dashed line) confidence levels calculated assuming Wilks' theorem are shown in gray for \citet{IceCube:2025arXiv} and in red for the analysis of IceTracks-DR2 data performed with SkyLLH.
  }
  \label{fig:ngc1068_lh_contour}
\end{figure}

NGC 1068 coincides with the most significant point-like source of neutrinos to date. To further test this data release, we perform a likelihood scan around the best-fit parameters and compute the 1$\sigma$ and 2$\sigma$ confidence intervals according to Wilks' theorem~\citep{WilksTheorem}. The comparison of the likelihood landscapes in Figure~\ref{fig:ngc1068_lh_contour} shows that the two overlap closely and exhibit the same shape as a function of the two parameters.

Finally, we evaluate the neutrino flux during the flare from the direction of \txs~\citep{IceCube:2018Science_flare} and compare the result to the time-dependent catalog analysis reported in \autoref{sec:tim-dep-catalog}. Because a full box-shaped flare search is not currently implemented in SkyLLH, we fix the time window to the best-fit parameters reported in \autoref{tab:tim-dep-catalog} and fit an unbroken power-law spectrum, leaving the number of signal events and the spectral index free. We obtain $\hat n_{\mathrm{s}}=12.68$ and $\hat\gamma=2.25$, values that are essentially identical to those reported in \autoref{tab:tim-dep-catalog} for the analysis of the same data using internal IceCube tools.


\section{Conclusion \& Outlook}
In this paper, we have presented IceTracks-DR2, the most comprehensive IceCube public data release to date, comprising 14 years of neutrino candidates observed between April 6, 2008 and May 23, 2022. This work supersedes IceTracks-DR1 \citep{IceCube:2021arXiv}, as the updated event sample presented here includes four additional years of data in addition to improvements in detector calibration and event processing (``Pass2" processing), leading to more accurate event filtering and reconstruction. 

Several generic point source search analyses were performed on the data using internal IceCube collaboration tools prior to releasing the data to the public. These include time-integrated and time-dependent all-sky and catalog searches, as well as a search for emission from the galactic plane. Documentation associated with reconstructing parts of these analyses using the contents of this data release and publicly available software tools were also shown in this paper. 

IceTracks-DR2 represents the largest and most comprehensive publicly available sample of TeV-PeV neutrino candidates available to date. With the advent of new astrophysical neutrino observatories in the near future ~\citep{Agostini_2020, km3net_loi}, the authors hope that this data release can serve as a template for future public release of data from similar neutrino telescopes worldwide.  

\bigskip

\textbf{Acknowledgments} The authors gratefully acknowledge the support from the following agencies and institutions: USA – U.S. National Science Foundation-Office of Polar Programs, U.S. National Science Foundation-Physics Division, U.S. National Science Foundation-EPSCoR, U.S. National Science Foundation-Office of Advanced Cyberinfrastructure, Wisconsin Alumni Research Foundation, Center for High Throughput Computing (CHTC) at the University of Wisconsin–Madison, Open Science Grid (OSG), Partnership to Advance Throughput Computing (PATh), Advanced Cyberinfrastructure Coordination Ecosystem: Services \& Support (ACCESS), Frontera and Ranch computing project at the Texas Advanced Computing Center, U.S. Department of Energy-National Energy Research Scientific Computing Center, Particle astrophysics research computing center at the University of Maryland, Institute for Cyber-Enabled Research at Michigan State University, Astroparticle physics computational facility at Marquette University, NVIDIA Corporation, and Google Cloud Platform; Belgium – Funds for Scientific Research (FRS-FNRS and FWO), FWO Odysseus and Big Science programmes, and Belgian Federal Science Policy Office (Belspo); Germany – Bundesministerium f\"{u}r Forschung, Technologie und Raumfahrt (BMFTR), Deutsche Forschungsgemeinschaft (DFG), Helmholtz Alliance for Astroparticle Physics (HAP), Initiative and Networking Fund of the Helmholtz Association, Deutsches Elektronen Synchrotron (DESY), and High Performance Computing cluster of the RWTH Aachen; Sweden – Swedish Research Council, Swedish Polar Research Secretariat, Swedish National Infrastructure for Computing (SNIC), and Knut and Alice Wallenberg Foundation; European Union – EGI Advanced Computing for research; Australia – Australian Research Council; Canada – Natural Sciences and Engineering Research Council of Canada, Calcul Québec, Compute Ontario, Canada Foundation for Innovation, WestGrid, and Digital Research Alliance of Canada; Denmark – Villum Fonden, Carlsberg Foundation, and European Commission; New Zealand – Marsden Fund; Japan – Japan Society for Promotion of Science (JSPS) and Institute for Global Prominent Research (IGPR) of Chiba University; Korea – National Research Foundation of Korea (NRF); Switzerland – Swiss National Science Foundation (SNSF).

\bibliography{references}{}
\bibliographystyle{aasjournalv7}

\end{document}